# Automatic Authoring of Physical and Perceptual/Affective Motion Effects for Virtual Reality

Jiwan Lee[1] and Seungmoon Choi[1]

[1] Dept. of Computer Science and Engineering, POSTECH, Pohang, South Korea

(Email: choism@postech.ac.kr)

**Abstract ---** **This video demo is about automatic authoring of various motion effects that are provided with audiovisual content to improve user experiences. Traditionally, motion effects have been used for simulators, e.g., flight simulators for pilots and astronauts, to present physically accurate vestibular feedback. At present, we have greatly wider use of motion effects for entertainment purposes, such as 4D rides in amusement parks and even shopping malls, 4D films in theaters, and relative new virtual reality games with head-mounted displays and personal motion platforms. However, the production of motion effects is done solely by manual authoring or coding, and this costly process prevents the faster and wider dissemination of 4D content. It is imperative to facilitate motion effect production by providing automatic synthesis algorithms. This demo introduces nine different automatic synthesis algorithms for motion effects and a recorded demonstration of each. All of these have been validated to deliver a reasonably competitive user experience in their respective studies.**

## 1 INTRODUCTION

Motion effects are indispensable for improving 4D experiences in highly interactive applications, such as amusement parks, 4D theaters, and virtual reality games. However, the production of motion effects is done solely by manual authoring or coding, and this costly process prevents the faster and wider dissemination of 4D content. It is imperative to facilitate motion effect production by providing automatic synthesis algorithms.

Motion effects are based on the sense of movement and balance perceived by the human vestibular system. As such, motion platforms generate physical stimuli in terms of linear accelerations and angular velocities, which can be represented by six independent variables, i.e., six degrees of freedom (DoFs). However, commercial motion platforms often provide fewer numbers of DoFs (Figure 1), and the most common are 2, 3, and 6. Moreover, the workspace of a motion platform is inherently limited, so it cannot provide a translational motion effect to one direction for a long time. Hence, designing effective motion effects is a challenging task, which must satisfy both of the goals pertaining to user experiences and the two requirements caused by the physical form factor. For this reason, all motion effects are designed and tested *manually and subjectively* using an in-house authoring tool. Consequently, motion effects authoring is very labor-intensive and costly in spite of the greatly wider dissemination of motion platforms in theme parks, 4D theaters, 4D rides, and VR games, as well as even personal ones. Further, the quality of motion effects is not evaluated using objective terms. These problems can be resolved or lessened by providing automated functions for authoring the output of which have been rigorously tested in terms of user perception and experiences. This is one of the must-to-overcome issues for more frequent and wider adoption of high-quality VR/4D content in a variety of applications.

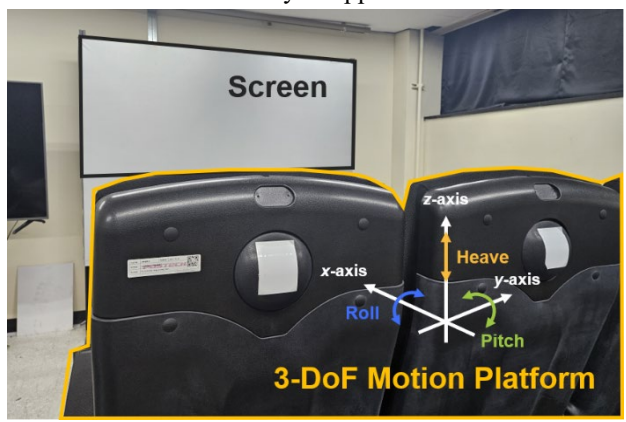

Figure 1. Environmental setup. Users can experience a video scene with motion effects. This motion chair supports the 3-DoF motion of roll, pitch, and heave.

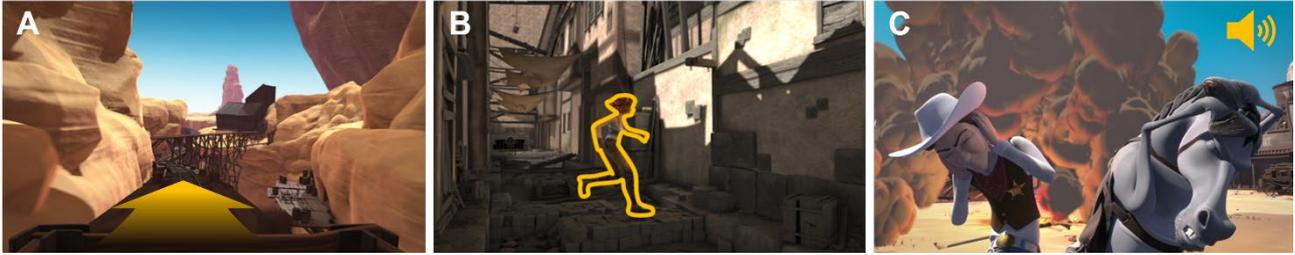

Figure 2. Three representative classes of motion effects. A: Camera motion, B: object motion, and C: Sound effects. These are the common authoring basis for motion effects in video used by 4D effect designers.

## 2 MOTION SYNTHESIS ALGORITHMS

This demo presents a series of nine different automatic authoring algorithms for motion effects, all of which have been published in reputable journals or conferences. Each algorithm has been validated to deliver a user experience that is competitive with manual work, while also generating motion effects in a reasonable amount of time. Refer to the respective studies for further details. These algorithms are presented in a video format instead of an on-site demonstration due to the heavy equipment, which is practically unmanageable for transport.

Motion effects capture and mimic 4D designers' common authoring methods, such as extracting visual motion or auditory information from videos. Motion effect authoring algorithms are classified according to the main authoring subject in the videos, such as cameras, objects, and sounds (Figure 2). Studies [1,7] focus on camera-based motion effects, studies [2,3,4,5] on object-based motion effects, while studies [6,8] consider motion effects based on two different bases.

### 2.1 Camera motion effects on first-person videos

This paper [1] introduces an initial algorithm for automatically generating camera-based motion effects. It generates the effects by estimating the camera position of a first-person perspective video. The estimated motion is expressed as a motion chair movement in a limited work space through a washout filter.

### 2.2 Object motion effects on static rigid bodies

This paper [2] calculates movement using 3D position and orientation information of static rigid objects and generates motion effects considering vestibular perception based on model predictive control. To create motion effects, the concept of a motion proxy is proposed, which expresses multiple motion information of an object as a single point.

### 2.3 Object motion effects on dynamic articulated bodies

This paper [3] proposes several motion proxy calculation methods based on joint and connection information of a dynamic multi-jointed body object (e.g., a moving person). Cases with multiple joints are also studied.

### 2.4 Object motion effects for general videos

This paper [4] generates object motion effects even for general images where object motion information is unknown. Motion effects are created by dividing the image into meaningful objects and estimating the 3D movement of the objects through scene flow estimation.

### 2.5 Object motion effects for human dancing

This paper [5] studies generating motion effects for human dancing. Caution is needed when human dance movements are compressed into a single movement proxy, such as the variety of movements for each upper and lower body. The quality of movement effect creation is improved by dividing body parts into consideration of dance movements and expressing each with an appropriate degree of freedom of movement.

### 2.6 Sound motion effects for shooting games

The key to creating motion effects in first-person gun games is real-time creation for e-sports viewing, etc. This paper [6] automatically detects the gunshot sound and calculates the motion effect representing the gun's recoil in real time. In addition, the speed of creating camera motion effects [1] is increased to near real-time level.

### 2.7 Perceptual motion effects for rough roads

It is difficult to express the road surface in the video at a high level through camera movement alone [1]. This study [7] generates fine motion effects by extracting roughness based on images similar to road surfaces and reflecting it.

### 2.8 Data-driven human gait motion effects

This paper [8] proposes a framework that generates various walking sensations as motion effects based on real data of human walking, and develop a generation algorithm that expresses motion effects in a low-dimensional space through principal component analysis and induces the target walking sensation.

### 2.9 Merging camera and object motion effects

This paper [9] covers several ways to merge various types of motion effects. Since each of them expresses the authoring subject well in the video, they try to merge them so as not to interfere with each other.

ACKNOWLEDGEMENT

This work was supported in part by Samsung Research Funding & Incubation Center (No. SRFC-IT1802-05), by the Mid-Career Researcher Program of the National Research Foundation of Korea (No. 2022R1 A2C2091161), and by the Pioneer Research Center Program through the National Research Foundation of Korea funded by the Ministry of Science, ICT & Future Planning (RS-2024-00451947).